\documentstyle[12pt,aaspp4]{article}
\begin{document}
\tighten
\def\etal{{\it et al.\/}}
\def\cf{{\it cf.\/}}
\def\ie{{\it i.e.\/}}
\def\eg{{\it e.g.\/}}

\title{A gamma ray burst model with small baryon contamination}
\author{{\bf Mario Vietri$^1$ and Luigi Stella$^2$}}
\affil{$^1$Universit\`a di Roma 3, Via della Vasca Navale 84, 00147 Roma, \\
Italy, E-mail: vietri@corelli.fis.uniroma3.it \\
$^2$ Osservatorio Astronomico di Roma, 00040 Monte Porzio Catone (Roma) \\
Italy, E-mail: stella@heads.mporzio.astro.it\\
Affiliated to I.C.R.A.}

\begin{abstract}
We present a scenario (SupraNova) for the formation of GRBs occurring when a 
supramassive neutron star (SMNS) loses so much angular momentum that centrifugal
support against self--gravity becomes impossible, and the star implodes to a 
black hole. This may be the baryon--cleanest environment proposed so far, 
because the SN explosion in which the SMNS formed swept the medium 
surrounding the remnant, and the quickly spinning remnant loses energy through 
magnetic dipole radiation at a rate exceeding its Eddington luminosity by
some four orders of magnitude. The implosion is adiabatic because 
neutrinos have short mean free paths, and silent, given the prompt 
collapse of the polar caps. However, a mass $M_l \approx 0.1\; M_\odot$ in
the equatorial belt can easily reach centrifugal equilibrium. The mechanism
of energy extraction is via the conversion of the Poynting flux (due to the 
large--scale magnetic field locked into the minitorus) into a magnetized 
relativistic wind.  Occasionally this model will produce quickly decaying, or 
non--detectable afterglows.  
\end{abstract}


\section{Introduction}

A successful model for the progenitors of gamma ray bursts (GRBs) must make 
contact with the fireball model (Rees and M\'esz\'aros 1992), which so 
successfully predicted the discovery of afterglows in the radio (Paczy\'nski 
and Rhoads 1993), in the optical (M\'esz\'aros and Rees 1997a, Katz 1994), in 
the X--ray band (Vietri 1997), and explained radio flares (Goodman 1997). 
While some features of afterglows remain at present unexplained,
yet there can be little doubt that the overall energetics, time-scales,
power--law behaviour, and appearance of longer wavelength radiation 
at later times are all nicely accounted for. 

The crucial point of the fireball model is that it channels the large explosion 
energy ($> 10^{53}\; erg$, barring beaming) in a locale with small baryon 
contamination ($\approx 10^{-4}\; M_\odot$, Rees and M\'esz\'aros 1992). From 
this point of view, hypernova explosions (Woosley 1993, Paczy\'nski 1998) are 
problematic: massive stars are well known to have powerful winds 
which may carry away upward of $10\; M_\odot$, thusly baryon--contaminating the 
environment in which the GRB is eventually to go off. Obviously, a very thin
beam may be a way out of this conundrum, but this begins to look  unlikely, 
given the currently available observational evidence (Grindlay 1998). Also 
scenarios involving accretion--induced collapse of white dwarfs (Usov 1992, 
Blackman, Yi, Field 1996, Yi and Blackman 1997) are troublesome, both because 
of the large amounts of baryons lying about, and because of the long timescales 
required to accrete sizeable amounts of mass. 

Given the considerable specific angular momenta involved, 
a plausibly baryon--clean model is the merger of neutron star/neutron star,
or neutron star/black holes binaries (Narayan, Paczy\'nski and Piran 1992)
which is expected to take place outside the environment where the binary formed
(Bloom, Sigurdsson and Pols 1998). However, at least some bursts (though 
not all, Wijers and Galama 1998) seem to arise 
within dense environments suggesting star forming regions: such is the case of 
GRB 970111 (Feroci \etal, 1998), GRB 970828 (Murakami \etal, 1997, Groot \etal, 
1998a) and GRB 980326 (Groot \etal, 1998b). Since it is possible that GRBs 
arise from several distinct sources (M\'esz\'aros, Rees and Wijers 1998b),
we propose here a different model, where the initial 
explosion is triggered by the gravitational implosion to a black hole of a 
supramassive neutron star: we call this model a SupraNova. 

In the following, we shall first discuss some properties of equilibrium
supramassive neutron stars, then the implosion to a back hole and the 
extraction of the energy that leads to the GRB. Reasons why this 
model remains baryon--clean will be discussed in Section 3, and in the last 
section we will discuss some observational consequences. 

\section{On supramassive neutron stars and their implosion}

All known equations of state for matter at nuclear densities allow the
construction of equilibrium models of supramassive neutron stars (SMNSs),
\ie\/ rotating sequences with constant baryon number but different angular
momenta; no member of this sequence may be non--rotating because 
they have masses at infinity larger than the largest mass for static models
(for recent work on the topic see
Cook, Shapiro and Teukolsky 1994, and Salgado, Bonazzola, Gourgoulhon and
Haensel 1994). Equilibrium sequences for models with baryon number 
exceeding a critical ${\cal B}_c$ above which no static equilibria are 
possible, start out exactly at break--up angular speed; depending upon
the equation of state, these models have masses in the range $M \approx
2-3.5\; M_\odot$, equatorial radii $R_{eq} \approx 11-18\; km$, and angular
velocities $\omega \approx 8,000-12,000 \; s^{-1}$. Though fast rotating, the 
total angular momentum of these models is not large: measured in units of the
critical angular momentum $J_c = GM^2/c$ of a black hole of the same mass
at infinity, they have $j \approx 0.6-0.78$. An important feature of these
models is that, as they lose angular momentum, through most of the models'
parameter space they speed up: they contract so as to reduce their moment of
inertia, and increase their angular velocity. For this reason, angular momentum 
loss through, for instance, magnetic dipole radiation actually speeds up.
As the losses mount, the models reach a point, having about half of the initial 
angular momentum, where they become secularly unstable to axisymmetric 
perturbations. The evolution of these models beyond this point has not been 
investigated yet, but the eventual fate of the star is sealed, since it cannot 
connect to a static, stable configuration: it will implode to a black hole. 

The above results hold for every equation of state, so that we regard them
as rather robust. Currently, there seems to be no obvious theoretical reason 
why these stars should not form
in the collapse of massive stars. A recent claim (Andersson 1998
%,Lindblom, Owen and Morsink 1998
) that a new type of r--modes leads to such a copious 
emission of gravitational waves, to make any fast--spinning neutron star slow 
down within a year, would seem to imply that SMNSs, which speed up as they lose
angular momentum, should collapse to a black hole faster than expected on the
basis of pure magnetic dipole radiation (Eq. \ref{tsd}, to be derived shortly). 
However, these computations only concern idealized neutron star models, and 
neglect proper, general relativistic treatment of viscosity, convection, and 
non--linear effects, all of which may slow down, or even damp this instability; 
they will be neglected henceforth. 
%In the next section, we shall show that, 
%at present, there ought to be just one SMNS within the Virgo cluster distance, 
%explaining why none has been detected yet. 

An approximate
estimate of the time it takes for this configuration to collapse is given
by assuming that the major source of angular momentum loss is through the usual
magnetic dipole radiation; the star's rotational energy is $E_{rot} = J \omega/2
$, and, since $\omega$ is roughly constant as the star loses angular momentum
passing through different stages of equilibrium
(Salgado \etal, 1994, Cook \etal, 1994), $\dot{E}_{rot} \approx \omega \dot{J}
/2$; equating this to the classical Pacini (1967) formula, one gets $\dot{J}$,
and the spin--down time to halve $J$, as required by the onset of the 
instability
\begin{equation}
\label{tsd}
t_{sd} \equiv \frac{J}{\dot{J}} = 10 \; yr \frac{j}{0.6} \left(\frac{M}
{3\;M_\odot}\right)^2 \left(\frac{15\;km}{R_{eq}}\right)^6 \left(\frac{10^4
\; s^{-1}}{\omega}\right)^4 \left(\frac{10^{12}\; G}{B}\right)^2 \;.
\end{equation}
Since this is much longer than the viscous timescale inside the NS, on which 
the configuration evolves after it has become secularly unstable toward the 
final collapse, $t_{sd}$ is a good estimate of the timescale between the first 
explosion (a normal SN where a SMNS is formed) and the implosion to a black 
hole. 

As the star loses angular momentum, the centrifugal force weakens until
support against self--gravity becomes impossible: the star implodes. 
The implosion is likely to be strongly non--homologous, the central densest 
regions forming a black hole first, with the outermost regions collapsing
later. To a first approximation, the implosion will be `silent', with most
matter being swallowed down the hole promptly, since the total angular 
momentum of the collapsing NS is subcritical ($j \approx 0.6-0.78$). 
In particular, it is likely that all mass lying close to the spin axis ends 
up in the hole, a point already made by M\'esz\'aros and Rees (1997b) and 
Paczy\'nski (1998). Furthermore, the neutrinos' mean free path inside the
NS is expected to be small, so the collapse will be, most likely, adiabatic.
Further baryon contamination must be small:
for instance, it is difficult to think of any outwardly propagating shock
capable of ejecting baryons, because of the lack of a hard surface and of 
energetic phenomena capable of releasing considerable counterpressures. Also,
for reasonable magnetic fields ($B < 10^{15}\; G$), any magnetic--related
phenomena using all the energy in the magnetic field could push at most 
$\approx 10^{-5}\; M_\odot$ in centrifugal equilibrium, all most likely in 
the equatorial plane. 
%Also, should any process manage to tap the energy in
%differential rotation, this too would most likely manage to shed a small 
%fraction of the mass in the equatorial plane. 

It is likely that this implosion is accompanied by the release of a copious 
amount of energy in gravitational waves; in fact, the known, axisymmetric 
instability of the equilibrium sequence is just secular (Friedman, Ipser
and Sorkin 1988); it is possible that this leads to a nearby, distinct 
equilibrium model which becomes dynamically unstable as it loses angular
momentum; the most likely source of instability is to a bar-formation mode 
($m=2$) which leads to release of gravitational waves (Chandrasekhar 1970). 

Not all mass will be immediately accreted, though: the outermost layers, in 
fact, have centrifugal accelerations only $\approx 10\%$ below the local 
gravitational
attraction (Salgado \etal, 1994). Thus, even if the pressure gradient were 
suddenly removed, these layers would just contract a bit (barring large 
energy losses due to shocks caused by the converging flow), and halt in 
centrifugal equilibrium at a radius $\approx 10\%$ smaller than the initial one.
We cannot compute $M_l$ at this stage, so we give only a suggestive argument 
showing that, possibly, $M_l \approx 0.1\; M_\odot$. 
Stellar dynamical simulations of self--gravitating galaxies in purely 
rotational equilibrium show that these objects are unstable to bar--forming 
$m = 2$ modes, a well--known result (Hohl 1971, Ostriker and Peebles 1973, 
Binney and Tremaine 1987). One may see these as energy conserving instabilities,
where stars may acquire some energy because the gravitational potential in 
which they move is not constant in time. In such systems, typically 
$\approx 1\%$ of the mass (originally, all stars on high angular momentum 
orbits) reaches orbits outside the main body of the resulting bar, becoming 
detached (but not unbound) from the galaxy. We may expect the total fraction of 
mass lost in hydrodynamic systems to be comparable to this, since here too 
parcels of fluid 
on high angular momentum orbits interact with the rest of the system mostly 
through the gravitational potential: in fact, according to the models of Salgado
\etal\/ (1994), the contribution of the pressure gradient to the support 
of the outer layers of the star is much smaller than the centrifugal one.
Moreover neutron stars differ from stellar dynamical systems 
because  of losses by gravitational waves, and possibly neutrinos, which
weaken  gravitational binding of the outermost layers: it is sufficient that
$\approx  10\%$ of the whole star mass at infinity ends up as GWs, for the
equatorial  belt to become rotationally--supported: this corresponds to
$0.3\; M_\odot$  radiated away, well within current estimates of GW emission
efficiencies in  non-axisymmetric collapses to black hole (Smarr 1979).
Neutrinos not trapped  inside the collapsing SMNS will also carry away
binding mass, or may provide  pressure pushing the equatorial belt into
centrifugal equilibrium. We thus  expect that the SMNS may typically shed a
few percent of its total mass to  centrifugal equilibrium. This implies $M_l
\ga 0.1 \; M_\odot$. 

The ensuing configuration is that of a largish black hole ($M \approx
2-3\; M_\odot$), spinning at subcritical speed ($j \approx 0.6$), surrounded
by a thin equatorial belt of matter of mass $M_l$ which could not be immediately
accreted because its angular momentum, supplemented by the factors discussed 
above, makes it rest in centrifugal equilibrium. The binding energy of the 
left--over mass $M_l$ is
\begin{equation}
%E_b = 5\times 10^{52}\; erg \frac{M_{SMNS}}{3\; M_\odot} \frac{M_l}{0.1\;
%M_\odot} \frac{15\; km}{R_{eq}}\;.
E_b = 3\times 10^{52}\; erg \frac{M_{SMNS}}{2\; M_\odot} \frac{M_l}{0.1\;
M_\odot} \frac{20\; km}{R_{eq}}\;.
\end{equation}
This binding energy must be radiated as the leftover accretes onto the black
hole, and it is this energy release which powers the GRB proper, not the
gravitational collapse {\it per se}. Even the
most powerful bursts (Kulkarni \etal, 1998) only require $M_l \approx 0.1 
\theta^2 M_\odot$, where $\theta^2/4$ is the beaming fraction. 

The leftover material will be threaded by the 
same, large--scale magnetic field lines that threaded it inside the NS; the
field will be amplified by shearing induced by differential rotation inside
the accreting torus. Assuming that the angular speed difference between the
front and back edges nearly equals $\omega$, a field of initial strength
$B \approx 10^{12}\; G$ will be amplified to $10^{15}\; G$ (\ie, well 
below equipartition with differential rotation, an upper limit which it is not
necessary to attain) in $\approx 10$ differential rotations; the amplification 
timescale is $t_a 
\approx 0.01\; s$, comparable to the bursts' rise--times. At this point 
the torus will radiate at the rate $\dot{E} \approx - B^2 \omega^4 r^6/6 c^3 =
10^{50}\; erg\; s^{-1}$, enough to power a burst. The details of the conversion 
of Poynting flux into a magnetized relativistic wind have been discussed in 
both different and similar contexts by Usov (1992, 1994), M\'esz\'aros and Rees 
(1997b), Katz (1997). Alternatively, one may invoke the Blandford--Znajek
(1977) mechanism to extract the much larger hole's rotational energy
(M\'esz\'aros, Rees and Wijers 1998b).

%This mechanism, also invoked by Paczy\'nski (1998), and originally suggested by
%Lovelace (1976) in a different context, takes place whenever the field is 
%severed from the black hole, and provides the viscosity necessary to remove the
%excess angular momentum from the torus. Should instead the field be dragged
%towards the black hole, then the Blandford--Znajek (1977) mechanism for the 
%extraction of the much larger hole rotational energy would become available. 
%Either way, a simple mechanism seems to exist to extract a large amount of 
%energy in a relatively baryon free environment.

The energy release rate computed above of course exceeds the Eddington 
luminosity of the minitorus by more than $10$ orders of magnitude. The energy
deposition occurs outside the minitorus, but it is of course quite likely 
that some mass is blown off the torus, for instance by photons that hit it 
and heat it. This baryon outflow will certainly contaminate the zones of the 
relativistic wind that are most distant from the spin axis, because
we do not expect any efficient angular momentum exchange in the process. 
Still, this implies that the outflow may be non--isotropic, and moderately 
beamed. The situation that we envision is similar to that described by
M\'esz\'aros and Rees (1997b), with a smoothly tapering beam. 

\section{A clean environment}

The volume immediately surrounding the SMNS will be vacated of baryons by
two effects. First, the SN explosion accompanying the SMNS's formation will
sweep away ISM baryons. We can gauge the importance of this effect by
considering the fact that most of the mass around the Crab pulsar, the only
plerionic remnant for which relevant data are available, is well--known
to be located in the filaments, which amount to no more than $\sim 5 M_\odot$ 
(Fesen, Shull and Hurford 1997), within about $3$ 
arcminutes. At a canonical distance of $2\; kpc$, this corresponds to a present
average baryon number density $n \approx 10\; cm^{-3}$. Assuming uniform 
expansion in the period from $t_{sd}$ to the Crab's current age, implies
that, at $t_{sd}$, the average baryon density was $n \approx 10^7\;
cm^{-3}$. Within a distance $D=10^{15}\; cm$, more than enough to guarantee
millisecond variability (which requires $D\approx 10^{12}\; cm$, Rees and 
M\'esz\`aros 1994) this corresponds to a total baryon mass $M \approx 3\times 
10^{-5} \; M_\odot$, below $10^{-4} M_\odot$, the upper limit to get a 
significantly relativistic expansion. We regard the above estimate as an
upper limit because the Crab is well--known to have an unusually low expansion 
velocity (a full factor of $3$ below other Type II SNRs, Fesen, Shull and 
Hurford 1997) and  because it considers the mass in the filaments as if it were 
uniformly distributed over the whole volume rather than concentrated in the 
observed thin shell. 

The second effect is due to the large energy release by the magnetic dipole 
rotation after the explosion, 
\begin{equation}
\label{eddington}
\dot{E} = -\frac{B^2 R^6 \omega^4 \sin^2\alpha}{6 c^3} = 
-3\times 10^{43}\; erg\; s^{-1} \left(\frac{B}{10^{12}\; G}\right)^2 
\left(\frac{R}{15\; km}\right)^6 \left(\frac{\omega}{10^4\;s^{-1}}\right)^4
\sin^2\alpha\;;
\end{equation}
since a large fraction of this energy will be converted to photons, the ensuing 
luminosity in fact exceeds the Eddington luminosity for the $\approx 3 M_\odot$ 
NS by about 4 orders of magnitude. This is of course identical to the 
well--known plerion model for SN remnants (Ostriker and Gunn 1971, Reynolds
and Chevalier 1984), except that the total energy released ($\approx 10^{53}\;
erg$) is much larger; for this reason we cannot directly apply these
models to our case, except to notice that its effect will surely be of 
lowering the naive baryon mass estimate obtained above ($\approx 3\times 10^{-5}
M_\odot$). 

\section{Observational implications}

We now discuss observational signatures of the occurrence of this model. It 
shares with all models for which GRBs are generated when young, massive
stars die a possibly exotic death, all predictions that associate GRBs with 
star forming regions. In particular, GRBs' afterglows should be localized 
within star forming galaxies, and their redshift distributions ought to match 
closely the history of star formation in the Universe (Madau \etal, 1996, 
Lilly \etal, 1996). Furthermore, the optical afterglow may be at times
absorbed  by dust present in the star--forming regions, and the X--ray one
may show  absorption by large equivalent column depths by neutral elements.
In fact, in the $t_{sd}$ years elapsed 
since the first explosion, the SN shock front will have reached a distance
$R_s = 10^{18}\; cm\; (v_{exp}/3\times10^9\; cm\; s^{-1})(t_{sd}/10\; yr)$.
The total column depth of baryons, at distance $R_s$, is 
$N_b = 10^{21}\; cm^{-2} (M_{ej}/10\; M_\odot) (R_s/10^{18}\; cm)^{-2}$
with $M_{ej}$ the ejecta mass; this column depth may perhaps be observable
(Murakami \etal, 1997). 
Thus we may sometimes expect to see the ionization edges discussed 
by M\'esz\'aros and Rees (1998a,b) due to dense, stationary outlying material. 

We argued above that in its maxi--Crab phase the SMNS will vacate of baryons 
a sizeable cavity around itself. Outside this cavity, however, there will 
be an outwardly increasing density gradient, a well--known property of Sedov 
solutions. In fact, matter behind the SN shock is subsonic with respect to the
outwardly propagating shock, so that it must be in approximate pressure 
equilibrium with previously shocked material. However, since the shock is
decelerating, just--shocked material will be at lower temperature than 
material shocked long before and, in pressure equilibrium, this implies higher 
densities. The exact Sedov solutions yield for the innermost regions $n_{ISM}
\propto r^d$  where $d = 3/(\gamma-1) = 9/2$ (Landau and Lifshitz 1979).
If the time between the SN explosion and the burst is rather short, it may
happen that the collision between different shells, which gives rise to the 
burst proper (Paczy\'nski and Xu 1994), takes place in a region which the 
maxi--Crab effect has not managed to empty of baryons, yet. In this case, 
the afterglow will propagate into the Sedov solution, \ie, inside an outwardly
increasing density gradient. 
In the afterglow, the time $t_{nr}$ of the transition to the non--relativistic 
expansion regime is reasonably sensitive to the density gradient; M\'esz\'aros, 
Rees and Wijers (1998a) find $t_{nr} \approx t_\gamma \Gamma^q$, where 
$t_\gamma$ is the duration the burst in the $\gamma$--ray band, $\Gamma \approx
100$ is the initial Lorenz factor of the ejecta, and $q = (8+2d)/(3+d)$, for
adiabatic expansion (notice that we defined as $-d$ what they call $+d$). 
For expansion in a wind, $d= -2$, and thus $q=4$, while for expansion into
an adiabatic Sedov solution, $d= 9/2$, $q=2.2$, so that the transition to
the non--relativistic regime is anticipated by up to four orders of magnitude
in time.  For $t_\gamma = 1\; s, q=2.2$, $t_{nr} \approx 7$ hours. After 
$t_{nr}$, the luminosity steepens considerably, to $t^{-2.4}$ or steeper 
(Wijers, M\'esz\'aros and Rees 1997). This provides thus a natural explanation
for those cases where either no afterglow is seen (in the hard X--ray GRB 
970111, Feroci \etal, 1998; in the optical GRB 970828, Groot \etal, 1998a), 
or it is seen to decrease more steeply than the other afterglows (GRB 980326, 
Groot \etal, 1998b). 

The formation rate of SMNSs in an $L_\star$--galaxy, $1/t_{SMNS}$, is of course 
expected to be of order of, but less than, the rate of formation for pulsars, 
in our Galaxy currently about $1/t_p$, with $t_p \approx 100\; yr$. If the 
formation rate is around a factor of $100$ lower than this, 
as suggested by rough considerations on the range of mass and angular 
momentum over which SMNSs form, there is ample room to 
account for all observed GRBs, including a moderate degree of beaming. 
It may prove interesting to search for such 
objects. 
The fraction of the time between each birth through which they are active in
the maxi--Crab phase is $t_{sd}/t_{SMNS}$, so that there will be one $L_\star$
galaxy every $t_{SMNS}/t_{sd}$ containing an active one. Given the luminosity
density of the Universe ($j_\circ = 1.7\times 10^8 L_\odot \; Mpc^{-3}$ in the
V--band) and the typical luminosity of an $L_\star$--galaxy like our own, 
($L_\star = 10^{10}\; L_\odot$ in the V--band, Binney and Tremaine 1987)
we find that the closest such active maxi--Crab should lie at a distance
%\begin{equation}
%R = \left(\frac{3 L_\star t_{SMNS}}{4\pi j_\circ t_{sd}}\right)^{1/3} =
%52\; Mpc
%\end{equation}
$ R = (3 L_\star t_{SMNS}/4\pi j_\circ t_{sd})^{1/3} = 52\; Mpc$
where we used Eq. \ref{tsd} and $t_{SMNS} = 10^5\; yr$ which is necessary if 
SMNSs must account for GRBs. At this distance, the flux of Eq. \ref{eddington}
corresponds to $\approx 10^{-10}\; erg\; s^{-1}\; cm^{-2}$, pulsed with
frequency $\nu = \omega/2\pi \approx 1000-2000 \; Hz$. 
If the Crab pulsar is anything to go by, we expect an X--ray flux of 
$\approx 10^{-11}\; erg\; s^{-1}\; cm^{-2}$  (only a fraction of
which will be pulsed) from a pointlike source well offset from the 
center of the galaxy. A detection of the pulsations from such a source is
beyond the capabilities of Rossi XTE, but well within reach of very 
large area X-ray telescopes such as XMM and, especially, Constellation. 

In summary, we presented a SupraNova model for gamma--ray bursts, \ie, a
model with a two--step collapse to a black hole not induced by 
accretion. The advantages of this model lie in its ability to keep the
environment baryon--free, both because of the SN explosion that goes with the
first collapse, which cleans up the surroundings, and because of the clean, 
silent, collapse of the NS once centrifugal support weakens critically. Second, 
we discussed a rotating magnetic dipole radiating much energy through its large 
rotation rate, which injects into the NS's surroundings
a luminosity exceeding Eddington's by four orders of magnitude, and which 
speeds up as a consequence of energy losses. This too
contributes to keeping the NS's environment baryon--clean. 
A rough estimate suggests that the total environment pollution by baryons
may be in the range of $\approx 10^{-4}\; M_\odot$.

\end{document}